\documentclass{aa}  

\usepackage{subfigure}
\usepackage{graphicx}
\usepackage{lineno}
\usepackage{multirow}
\usepackage{natbib}
\bibpunct{(}{)}{;}{a}{}{,} 
\usepackage{txfonts}
\begin{document} 


   \title{The primordial environment of super massive black holes: \\large scale galaxy overdensities around $z\sim6$ QSOs with LBT.}


  \author{L. Morselli
          \inst{1},
          M. Mignoli\inst{2},
          R. Gilli\inst{2},
	  C. Vignali\inst{3,2},
         A. Comastri\inst{2},
	 E. Sani\inst{4},
         N. Cappelluti\inst{2},
	 G. Zamorani\inst{2},
	 M. Brusa\inst{3,2},
	S. Gallozzi\inst{5},
	E. Vanzella\inst{2}
         }
   \institute{Excellence Cluster Universe, Boltzmann str. 2, D-85748 Garching, Germany
  \email{laura.morselli@tum.de}
	\and INAF - Osservatorio Astronomico di Bologna, via Ranzani 1, 40127 Bologna, Italy
	\and Dipartimento di Fisica e Astronomia, Universit\`a degli Studi di Bologna, viale Berti Pichat 6/2, 40127 Bologna, Italy
	\and INAF – Osservatorio Astrofisico di Arcetri, largo E. Fermi 5, 50125, Firenze, Italy
	\and INAF - Osservatorio Astronomico di Roma, via di Frascati 33, I-00040 Monteporzio, Italy
             }

 \abstract
   {In the current model of structure formation, bright QSOs at $z\sim6$ are supposed to be hosted by the most massive dark matter halos collapsed at that time. The large 
scale structures in which these halos are embedded may extend up to 10 physical Mpc, and should be traced by overdensities of star-forming galaxies. 
To date, the search for such overdensities has been limited to scales of 1-2 physical Mpc around the QSO and did not produce coherent results.}
   {We aim at studying the environment of $z\sim6$ QSOs and verify whether they are associated to large scale overdensities of galaxies selected at the same redshift as 
$i$-band dropouts.}
   {With the wide-field ($\sim23'\times25'$) Large Binocular Camera (LBC) at the Large Binocular Telescope (LBT), we obtained deep $r$-, $i$- and $z$- band imaging of the
fields around four high-z QSOs, namely SDSS J1030+0524 (z = 6.28), SDSS J1148+5251
(z = 6.41), SDSS J1048+4637 (z = 6.20) and SDSS J1411+1217 (z = 5.95). 
Our photometric catalogues are based on source detection in the $z$-band image (5$\sigma$) and contain from
$\sim 2.3\times 10^4$ to $\sim 2.9\times 10^4$ objects, down to a 50\% completeness
limit of $z$=25.0-25.2 AB mag. We adopted color-color selections within the $i-z$ vs $r-z$ plane to identify samples of $i$-band dropouts at the QSO redshift and 
measure their relative abundance and spatial distribution in the four LBC fields, each covering $\sim8\times8$ physical Mpc at $z\sim6$. 
The same selection criteria were then applied to $z$-band selected sources in the $\sim$1 deg$^2$ wide-and-deep Subaru-XMM Newton Deep Survey (SXDS) to derive 
the expected number of dropouts over a blank LBC-sized field ($\sim$0.14 deg$^2$ after removing masked regions).}
   {The four observed QSO fields host a number of candidates larger than what is expected in a blank field. By defining as $i$-band dropouts objects with $z_{AB}<25$, $i-z>1.4$ and undetected in the $r$-band, 
we found 16, 10, 9, 12 dropouts in SDSS J1030+0524, SDSS J1148+5251, SDSS J1048+4637, and SDSS J1411+1217, respectively, whereas only 4.3
such objects are expected over a 0.14 deg$^2$ blank field.
This corresponds to overdensity significances of 3.3, 1.9, 1.7, 2.5$\sigma$, respectively, after accounting for cosmic variance and for the contamination by bluer 
objects in our dropout samples produced by photometric errors. By considering the total number of dropouts in the four LBC fields and comparing it with what is expected
in four blank fields of 0.14 deg$^2$ each, we find that high-z QSOs reside in overdense environments at the $3.7\sigma$ level. This is the first direct and unambiguous
measurement of the large scale structures around $z\sim6$ QSOs.
}
   {}

\titlerunning{Large scale galaxy overdensities around $z\sim6$ QSOs.}
\authorrunning{Morselli et al.}
   \maketitle
%

\section{Introduction}

Wide-area optical surveys like the Sloan Digital Sky Survey  \citep{f01a} and the Canada France High-z Quasar Survey \citep{wil1, wil2} have discovered a few tens of 
$5.7 < z < 6.5$ quasars \citep{f01a, f01b, f01c, f03} powered by accreting supermassive black holes (SMBHs)  with M$_{BH}$ > 10$^{8-9}$ M$_{\odot}$ \citep{kurk,der}. 
Wide-area infrared surveys, such as the UKIRT (United Kingdom Infrared Telescope) Infrared Deep Sky Survey  \cite[UKIDSS; see][]{law} and the public ESO-VISTA 
surveys \citep{arnab} are now moving to $z>6.5$ the redshift frontier for QSOs discovery. 
In particular,  three QSOs at 6.5< z < 7.0 have just been discovered by the VISTA Kilo-degree Infrared Galaxy survey (VIKING; see \citealt{ven}). The most distant QSO known, 
at  z = 7.08, was discovered two years ago by the UKIDSS Large Area Survey \citep{mortlock}. \cite{der1} studied the properties of these 
QSOs at z > 6.5 and found that they host $\sim10^9$ M$_{\odot}$ BHs. To explain the existence of such massive objects in place only 1 Gyr after the Big Bang, structure formation 
models assume that SMBHs are hosted in the most massive dark matter halos collapsed at that time \citep{vol}. A BH seed, born at z $\sim$ 20 as the consequence of the direct 
collapse of a gas cloud or as end-product of the first generation of stars, may experience major merging events that trigger gas accretion
and can become a 10$^9$ M$_{\odot}$ SMBH in a relatively short time, shorter than 1 Gyr.  If this were true, the fields around high redshift QSOs are expected to show galaxy 
overdensities, which should embed the progenitors of the most massive clusters of the Local Universe, with M > 10$^{14-15}$ M$_{\odot}$ \citep{spring}. 
This is in agreement with the average 
mass of  the  z = 0 descendants of the halos hosting $z\sim6$ QSOs, as found by recent simulations (albeit with significant, $\sigma$ = 0.36 dex, scatter; see \citealt{ang}). 

In the last decade, large efforts have been made to understand the properties of the $z\sim6$~QSO fields, which are extremely rare ($\sim$1 per comoving Gpc$^{3}$) and are 
supposed to be hosted by dark matter halos of $M_{h}$=$10^{13}$ M$_{\odot}$ \citep{f04}. \citet[hereafter S05]{stiav} found an excess of candidate Lyman Break Galaxies (LBGs) around the 
QSO SDSS J1030+0524 at z = 6.28 using ACS observations, that cover a field of 3'$\times$ 3' around the central QSO ($\sim$ 1$\times$1 physical Mpc). \cite{wil3} studied the field 
around three $z\sim6$~QSOs using the GMOS-North imaging spectrograph, that has a field of view of 5'$\times$5', corresponding to $\sim$1.8$\times$1.8 physical Mpc at $z\sim$6. 
They found no evidence for galaxy overdensity in all the three fields. \citet[hereafter K09]{kim} observed the fields around five $z\sim6$~QSOs with ACS and found two of them to be 
overdense, two underdense and a 
normo-dense one. Recently, \cite{ban} observed the field around the QSO ULAS J0203+0012 at z =5.72 with the FORS2 spectrograph, which has a FoV of 
6.8'$\times$6.8'$ (\sim$ 2.5$\times$2.5 physical Mpc). They found that the number of Lyman Alpha Emitters (LAEs) is consistent with what is expected in blank fields, 
and tried to interpret the lack of an 
overdensity with different scenarios, including also the possibility that the strong ionizing radiation from the QSO may prevent star formation in its vicinity, 
or that $z\sim6$~QSOs are not hosted in the most massive dark matter halos. The observational scenario is unclear even at lower redshifts: observations of QSOs at 2 < z < 5 
revealed galaxy overdensities around some of them \citep{dj,kas,can,swin,hus} and average-densities around some others \citep{fran}.  In addition, even the richest environments 
around them are as rich as other observed structures that do not host a SMBH \citep{hus}. 

Recent simulations by \cite{fanidak} have suggested that halos hosting luminous $z\sim6$~QSOs have masses around 10$^{11}$-10$^{12}$ M$_{\odot}$, i.e. at least a factor 
of 10 smaller than that suggested by \cite{vol}. As a consequence, they would evolve in structures that are not the most massive ones in the local Universe, and should not
be likely associated to galaxy overdensities at $z\sim6$.
However, the apparent lack of strong galaxy overdensities around high-z QSOs is not in contrast with them to be hosted in the peaks of the dark matter distribution. This is shown by the 
simulation of \cite{overz}, based on the Millennium Run \citep{spring1}.  They showed that galaxy overdensities can extend on scales of 10 physical Mpc or more from the central 
BHs, i.e. on scales that are larger than those observed using ACS or GMOS. This result puts new light in the study of the $z\sim6$~QSOs environment: to probe the typical density of the 
fields in which those extreme structures form, $wide$ $area$ and $deep$ observations are needed. In light of this, \cite{ut} carried out a study around the QSO CFHQS J2329-0301 at 
z = 6.43 using the wide FoV of the Supreme Cam (34'$\times$27') at the Subaru Telescope and they found a $\sim3\sigma$ overdensity of galaxies around it on scales of $\sim 5-6$ physical Mpc.\\
\indent In this work, we study the fields around four of the most distant QSOs known to date using wide area, deep images in the $r$-, $i$- and $z$- bands, obtained with the Large Binocular Cameras 
(LBC; \citealt{giall}) at the Large Binocular Telescope (LBT\footnote{\sl The LBT is an international collaboration among institutions in the United States, Italy, and Germany. LBT Corporation partners are: The University of Arizona on behalf of the Arizona university system; Istituto Nazionale di Astrofisica, Italy; LBT Beteiligungsgesellschaft, Germany, representing the Max-Planck Society, the Astrophysical Institute Potsdam, and Heidelberg University; The Ohio State University; and The Research Corporation, on behalf of The University of Notre Dame, University of Minnesota and University of Virginia.}) on Mount Graham in Arizona.
LBC is caracterized by a unique $etendue$, the product between collecting area and field of view, of 111 m$^2$$\times$ 0.16 deg$^2$ (considering both telescopes). This allows 
the study of $z\sim6$ overdensities on scales of 8 $\times$ 8 physical Mpc around the QSOs ($\sim56$ comoving Mpc), i.e. on areas at least a factor of 20 larger than those explored by previous studies with ACS os GMOS.

The paper is outlined as follows: in Section 2 the observation, data reduction and the criteria to select $z\sim6$ galaxies are described. In Section 3 the candidates in each field are shown. In Section 4 we discuss our results for the four observed fields, while in Section 5 we present our conclusions. \\
\indent We adopt a $\Lambda$CDM cosmology, with H$_0$ = 70 km s$^{-1}$ Mpc$^{-1}$, $\Omega_M$ = 0.3 and $\Omega_{\Lambda}$ = 0.7.


\section{Observations and data  analysis}
	\subsection{Observations}

We investigated the fields around  SDSS J1030+0524 (z = 6.28, Fan et al. 2001b), SDSS J1148+5251 (z = 6.41, Fan et al. 2003), SDSS J1048+4637 (z = 6.20, Fan et al. 2003) and SDSS J1411+1217 (z = 5.95, Fan et al. 2004). The four fields (hereafter J1030, J1148, J1048 and J1411) have been observed with the $z_{SDSS}$, $i_{SDSS}$ and $r_{SDSS}$ filters of  LBC.  Celestial coordinates, redshifts, BH masses \citep{der} and $z_{AB}$ magnitudes of the four $z\sim6$~QSOs are shown in Table~\ref{quasar}.  These QSOs have been chosen because they host some of the most massive black holes observed at $z\sim6$ (1.4 - 4.9 $\times$ 10$^9$ M$_{\odot}$, see Table~\ref{quasar}) and for observability reasons. 

\begin{table*}
\renewcommand{\tablename}{Tab.}
\caption{Celestial coordinates, redshift, BH mass and $z_{AB}$ magnitude of the four studied quasars (from \citealt{der}).}
\begin{center}
\begin{tabular}{cllcccc}
\hline
\hline

Field & RA (J2000) & Dec (J2000) &z & $M_{BH}$ &$ z_{SDSS}$  \\
 & & & & [$10^9M_{\odot}$] & [AB mag] & \\
\hline
J1030 & 10:30:27.1 & +05:24:55 & 6.28 & 3.2 & 20.0 \\
J1148 & 11:48:16.1 & +52:51:50 & 6.41 & 4.9 & 20.1\\
J1048 & 10:48:45.5 & +46:37:18 & 6.20 & 3.9 & 19.9  \\
J1411 & 14:11:11.3 & +12:17:37 & 5.95 & 1.2 & 19.6 \\
\hline
\label{quasar}
\end{tabular}
\end{center}
\end{table*}

The FoV of LBC is 23.6$\times$25.3 arcmin$^2$, which corresponds to $\sim$ 8.1$\times$8.7 physical Mpc at z $\sim$ 6. Each field is observed  using three hours of exposure, 1.5hr in $z_{SDSS}$ and 1.5 hr in $i_{SDSS}$ on the LBC-red 
channel, and simultaneously 3 hr in the $r_{SDSS}$ filter on the LBC-blue channel. This is done in order to reach AB sensitivity limits that are efficient in the colour selection of $z\sim6$ sources (see Sec.\ref{selection}).
 
Standard data reduction, i.e. flat fielding, sky subtraction, image alignment and stacking, was performed with the LBC pipeline. In addition, a mask with an exposure map has been used to mask low exposure time regions due to the 
dithering technique. A detailed description of the reduction procedure can be found in \cite{giall}. Finally,  visual inspection was needed to mask spikes due to saturation, bright stars haloes (due to reflections within the cameras) and artifacts. 

The seeing FWHM for each band image and for each QSO field is shown in Table~\ref{riass}. The seeing values have been computed from the brightness profile of reliable point-like objects. Table~\ref{riass} also shows the effective area of 
each field, after masking low S/N and contaminated regions, and the total number of detected objects per field.

\subsection{Object Detection}

Object detection has been carried out using the software SExtractor 2.3.2 \citep{bee}. The $z$-band images have been used as $master$ images in which object detection is performed. Several tests have been carried out to obtain the set 
of input parameters that correspond to the maximum number of real objects while keeping suspected false detections as few as possible. In particular, we considered an object detected if it is made of 9 connected pixels, each exceeding 
the local sky background by of a factor of 1.5 (for a total 4.5$\sigma$ detection). We then used SExtractor in dual mode to obtain the photometric measurements on the $i$ and $r$ images. We chose an aperture diameter of 7 pixels 
($\sim1.6$ arcsec) to compute the aperture magnitudes in order to get a large fraction of the flux from the object while minimizing the contamination from neighbouring sources. The four created catalogs (one per field) have typically 
$\sim2.5\times 10^4$ detected objects each. The field with the largest number of sources is J1030, which is observed in the $z$-band with the best seeing. 

The photometric zero points took into account the correction for dust extinction computed from \cite{schl}. We used the SExtractor MAG$\_$AUTO magnitudes\footnote{MAG$\_$AUTO is computed using an adaptive
elliptical aperture around every detected object, following \citep{kron80}.} to estimate the total magnitudes (e.g. $z_{TOT}$), because of their low
uncertainties and small systematic errors at faint magnitudes. Finally, we included an aperture correction term to optimise the estimate of the aperture magnitude for pointlike
objects since we needed to compensate for seeing variations between different fields/filters.

The completeness limit in total magnitude of the images, defined as the value of the magnitude at which the number of detected objects falls at 50$\%$ of the expected value, is 25.2 for the $z$ image of the J1030 field (the best seeing image, 
see  Table\ref{riass}), while it is 25.0 for the other three $z$ images. The $z$-band selected catalogs with associated $r$- and $i$-band photometry, as well as the reduced $r$, $i$, $z$-band images of the four fields
are publicly available.\footnote{\url{http://www.oabo.inaf.it/~LBTz6}}

The selection criteria of $z\sim6$ source candidates makes use of colors computed from aperture magnitudes corrected for seeing variations, therefore we also need to estimate the completeness limits of the aperture magnitude 
in the $i$ and $r$ bands. 
To compute these limits, we adopt a deeper 3$\sigma$ object detection, performed independently in the $i$ and $r$ images of the four fields. The $i$ and $r$ detected catalogues  show very similar photometric properties so, 
in order to achieve uniformity in the selection criteria, we adopted for all the four different fields the $i$ and $r$ magnitude aperture limits of 26.6 and 27.2, respectively.

\begin{figure}
  \resizebox{\hsize}{!}{\includegraphics{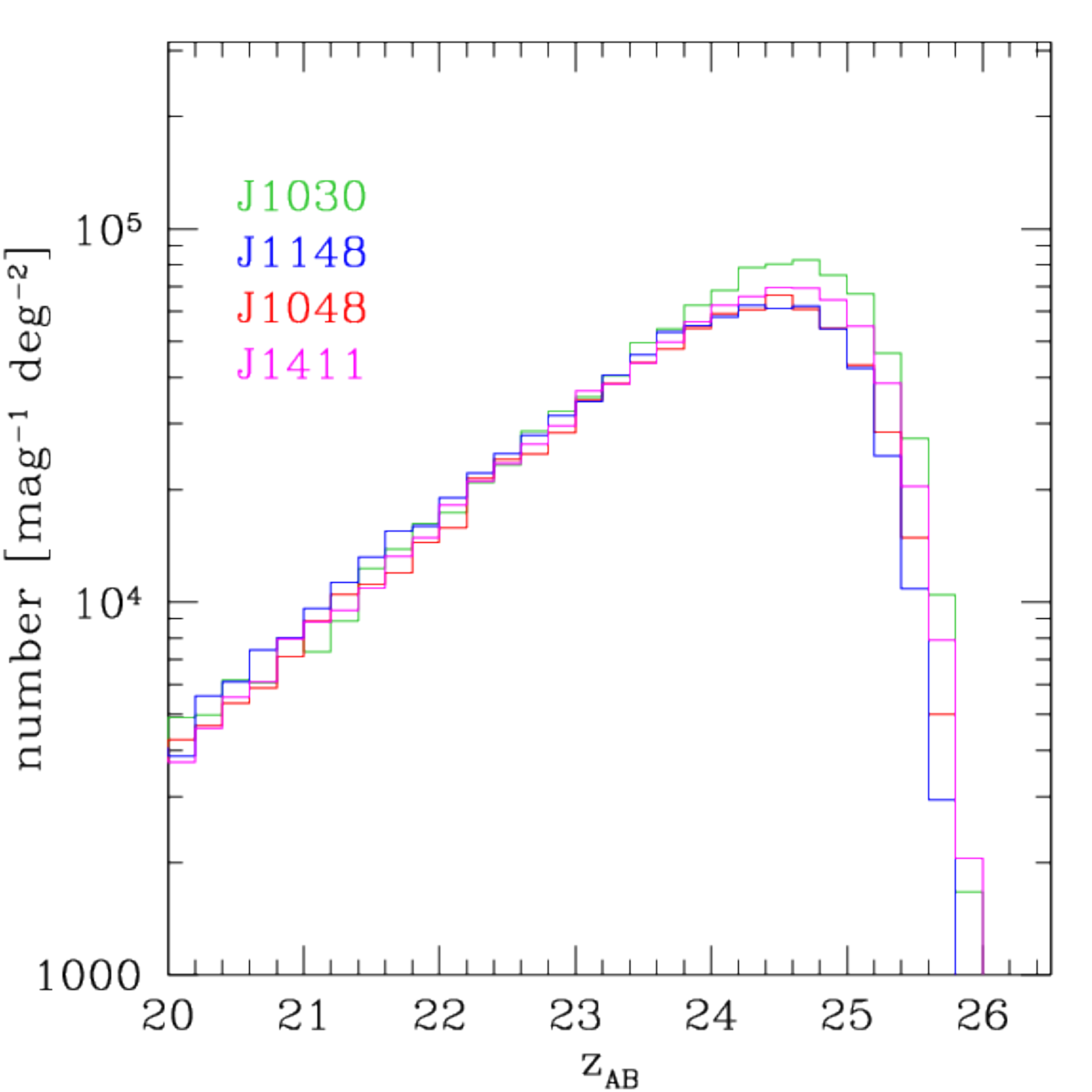}}
  \caption{Source number counts in the z-band filter in the four LBC fields. The green, blue, red and magenta histograms refer to J1030,  J1148, J1048 and J1411, respectively.}
  \label{counts}
\end{figure}

\indent To confirm the reliability of our photometric measurements, catalogs of SDSS sources in the same area of the four LBC fields have been downloaded from the SDSS archive and then cross-correlated with the LBC catalogs, using a 
correlation radius of 0.8 arcsec (this value has been chosen to minimize the number of contaminants). In particular, for point-like sources we compared the SDSS $psfMag$ with the total magnitude of LBC catalogues. 
The mean difference $\langle\Delta_{mag}\rangle$ between the two magnitudes in the three filters is small\footnote{In particular, the $\langle\Delta_{mag}\rangle$ = psfMag(SDSS) - MagTot(LBC) in the three filters is: 
-0.10 in $z$, +0.07 in $i$ and -0.17 in $r$. This implies an $i-z=0.17$ color shift between the two photometric systems.}, 
no offset is seen between the four LBC fields, and the $rms$ of the $\Delta_{mag}$ distribution in the three filters is less than 0.1, confirming the reliability and stability of the photometric measurements of this work.  Fig.\ref{counts} shows 
the number counts per deg$^2$ of sources in $z_{TOT}$ in the four fields.  The number counts in the four fields are in excellent agreement with each other.

\begin{table*}[htbp]
\renewcommand{\tablename}{Tab.}
\caption{Seeing (FWHM) in the $z$, $i$ and $r$ filters, effective area of the image, number of detected sources, 50$\%$ completeness limit in the $z$-band image,  
and number of primary ($(i-z)-\sigma_{(i-z)} >1.3$) and secondary (1.1 < $(i-z)-\sigma_{(i-z)}$ < 1.3) $z\sim6$ LBG candidates in the four fields.}
\begin{center}
\begin{tabular}{lccccccc}
\hline
\hline
& & & & & & & \\
Field & Filter     & Seeing {\footnotesize FWHM}    & Clean Area& Total Objects& Completeness Limit& Primary  & Secondary\\
 &                     &                      [arcsec]                  &       [deg$^2$]  & z$_{TOT}$ &                                &   candidates & candidates\\
\hline
\multirow{3}*{J1030} 	& $z$     	&   0.64 $\pm$ 0.02 	& \multirow{3}*{0.143}  & \multirow{3}*{29151}  & \multirow{3}*{25.2} & \multirow{3}*{14} & \multirow{3}*{10}	        \\
 				& $i$      	&   0.70 $\pm$ 0.02 	&   																					\\
 				& $r$      	&   0.74 $\pm$ 0.02 	& 																					 \\
\hline
\multirow{3}*{J1148} 	& $z$ 	&  0.79 $\pm$ 0.02 	& \multirow{3}*{0.144} & \multirow{3}*{24447}  & \multirow{3}*{25.0} &\multirow{3}*{8} & \multirow{3}*{3$^a$} 			\\
 				 & $i$ 	&  0.87 $\pm$ 0.02	& 																					 \\
 				 & $r$ 	&  0.93 $\pm$ 0.02 	& 																					\\
\hline

\multirow{3}*{J1048} 	& $z$ 	& 0.84 $\pm$ 0.01	& \multirow{3}*{0.143} & \multirow{3}*{24087}  & \multirow{3}*{25.0}  &\multirow{3}*{6} & \multirow{3}*{9} 			\\
 				 & $i$ 	& 0.87 $\pm$ 0.02	&																					 \\
 				 & $r$ 	& 1.04 $\pm$ 0.02	& 																					 \\
\hline

\multirow{3}*{J1411} 	& $z$ 	& 0.73 $\pm$ 0.01	&\multirow{3}*{0.145} & \multirow{3}*{26055}  & \multirow{3}*{25.0} &\multirow{3}*{11} & \multirow{3}*{8}			 \\
 				 & $i$ 	& 0.77 $\pm$ 0.01	& 																					\\
 				 & $r$ 	& 0.80 $\pm$ 0.01	&																					\\
\hline
\label{riass}
\end{tabular}

$^a$In the J1148 field an object that just barely misses our selection criteria for secondary $z\sim6$ LBG candidates (i.e. it is undetected in the $r-$band, has $z_{tot}=23.4$, but has $(i-z)-\sigma_{(i-z)} =1.06$), was proven
to be a $z=5.7$ AGN \citep{maha}.
\end{center}
\end{table*}

\subsection{Candidate Selection}
\label{selection}

At high redshift, galaxy spectra are characterized by a flux drop blueward of the rest-frame wavelength of the Ly$\alpha$ line ($\lambda_{Ly\alpha,rest}$ = 1216 $\AA$) as a 
consequence of the increasing amount of neutral hydrogen in the intergalactic medium. This drop is the so-called Gunn-Peterson through \citep{gp} and is routinely used to 
identify galaxies at high redshift. At $z\gtrsim5.9$, the observed wavelength of the Ly$\alpha$ line is at $\gtrsim$ 8400 $\AA$, i.e. the line moves in the $z_{SDSS}$ filter. 
As a consequence, the flux in the $i_{SDSS}$ filter is suppressed and the $z\sim6$ galaxies can be identified as $i$-band dropouts. In particular, \cite{dick}, \cite{beck} and 
other authors have shown that, using different spectral templates of mostly star forming galaxies, objects at z > 5.6 can be efficiently selected among those with $i-z>1.3$. 

The $z\sim6$ source candidates identified using the $i-z$ color can be affected by contamination from red objects at lower redshift. In particular, elliptical galaxies are 
characterised by a large Balmer break ($D_{4000}$, $\lambda_{rest}$ = 4000 $\AA$) in their spectra and, at z $\gtrsim1.1$, the $D_{4000}$ falls in the $z$ filter, 
causing the source to be detected as $i$-dropout. To reduce the contamination from elliptical galaxies it is possible to use the $r-z$ color (similarly to the ACS/HST $V-i$ color criterium
adopted by \citealt{beck} and \citealt{bouwens06}). In fact, the flux of elliptical galaxies 
blueward of the $D_{4000}$ does not drop to zero as the expected flux of $z\sim6$ galaxies beyond the redshifted Lyman limit. For this reason, $z\sim6$ galaxies have redder 
$r-z$ color than elliptical galaxies at lower redshift. Moreover, since the profile of elliptical galaxies at $z\sim1-2$ should be resolved in our sharp LBC images, we can use the 
morphology of the selected sources to estimate the contamination from elliptical interlopers in the sample of high redshift galaxies (see Section~4.1).

Cool dwarfs like L and T stars with temperatures in the range 700-2000 K, have intrinsic colors similar to those of $z\sim6$ galaxies. It is difficult to quantify the degree 
of contamination by late-type dwarfs in our
fields since there is no accurate knowledge of late-type star counts at different sky positions at the faint magnitudes we are probing. We limit ourselves to note that our LBC fields
are all at high Galactic latitude ($b>50$), and three of them point far away from the Galactic center (240<$l$<140). The only field that points towards the Galactic center is J1411, at $l$=359, which may then 
suffer from some stellar contamination more than the other fields. We note, however, that the precise knowledge of the stellar contamination is unlikely to bias our results significantly. 
Indeed, the Subaru XMM-Newton Deep Survey that will be used in Section 4.3 as a reference field to quantify galaxy overdensities, has Galactic coordinates comparable with those of at least three 
of our four fields, and should thus share a similar level of stellar contamination among $i-$band dropouts (see Section 4.3).


 \begin{figure*}
 \centering
 \includegraphics[scale=0.7]{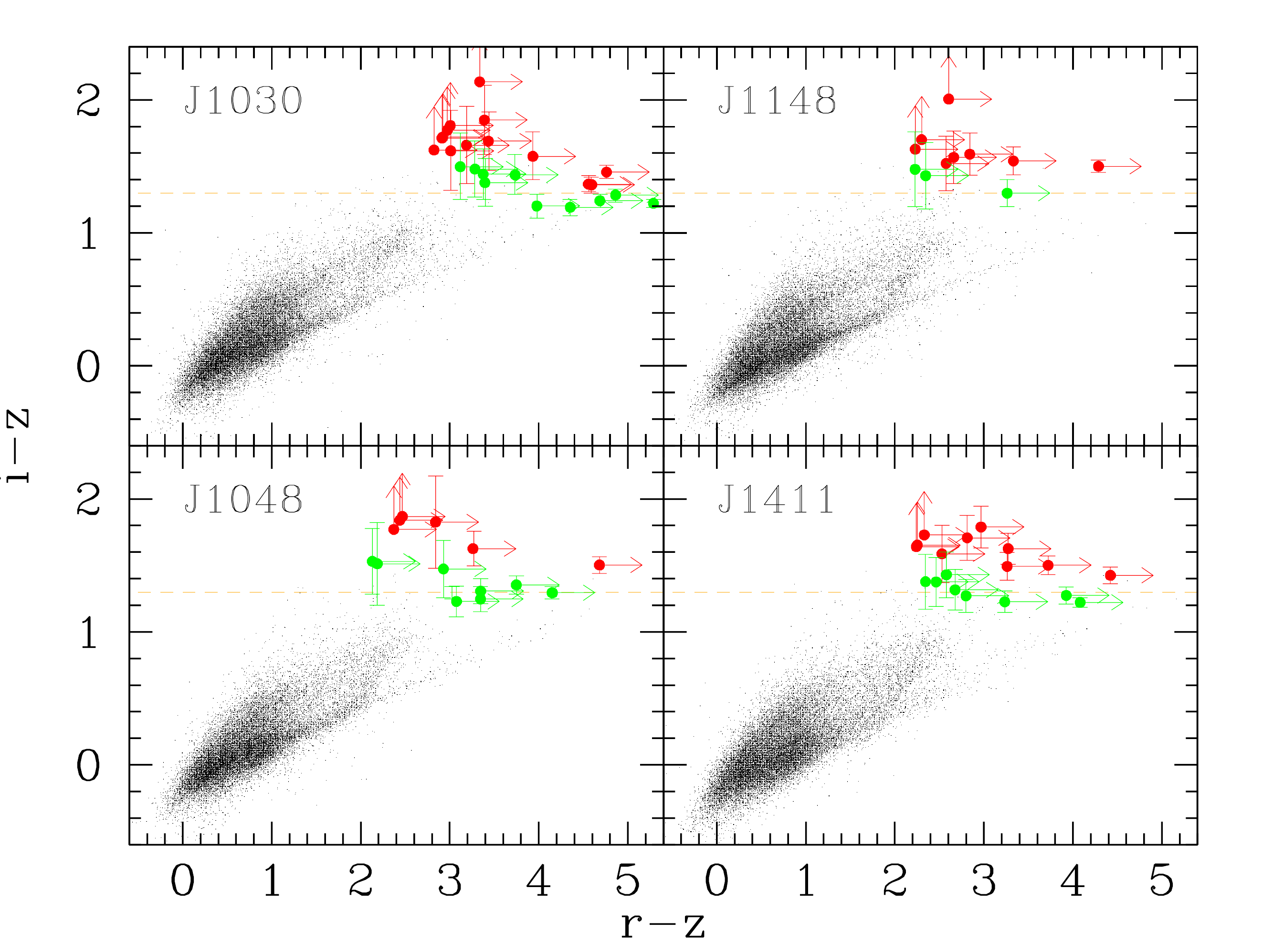}
\caption{$i-z$ versus $r-z$ diagram for the sources in the four QSO fields (small black dots). Primary and secondary $z\sim6$ galaxy candidates are marked with red and green dots, respectively.
The dashed line represents the adopted threshold of 1.3 in $(i-z) - \sigma_{i-z}$ used to select primary candidates. A primary candidate in the J1411 field (with $i-z>2.62$) falls outside the plot.
The small black dots above the dashed lines do not match all of our selection criteria (e.g. most of them are detected in the $r-$ band).}
\label{color}
 \end{figure*}

 \begin{figure*}
 \centering
   {\includegraphics[scale=0.36]{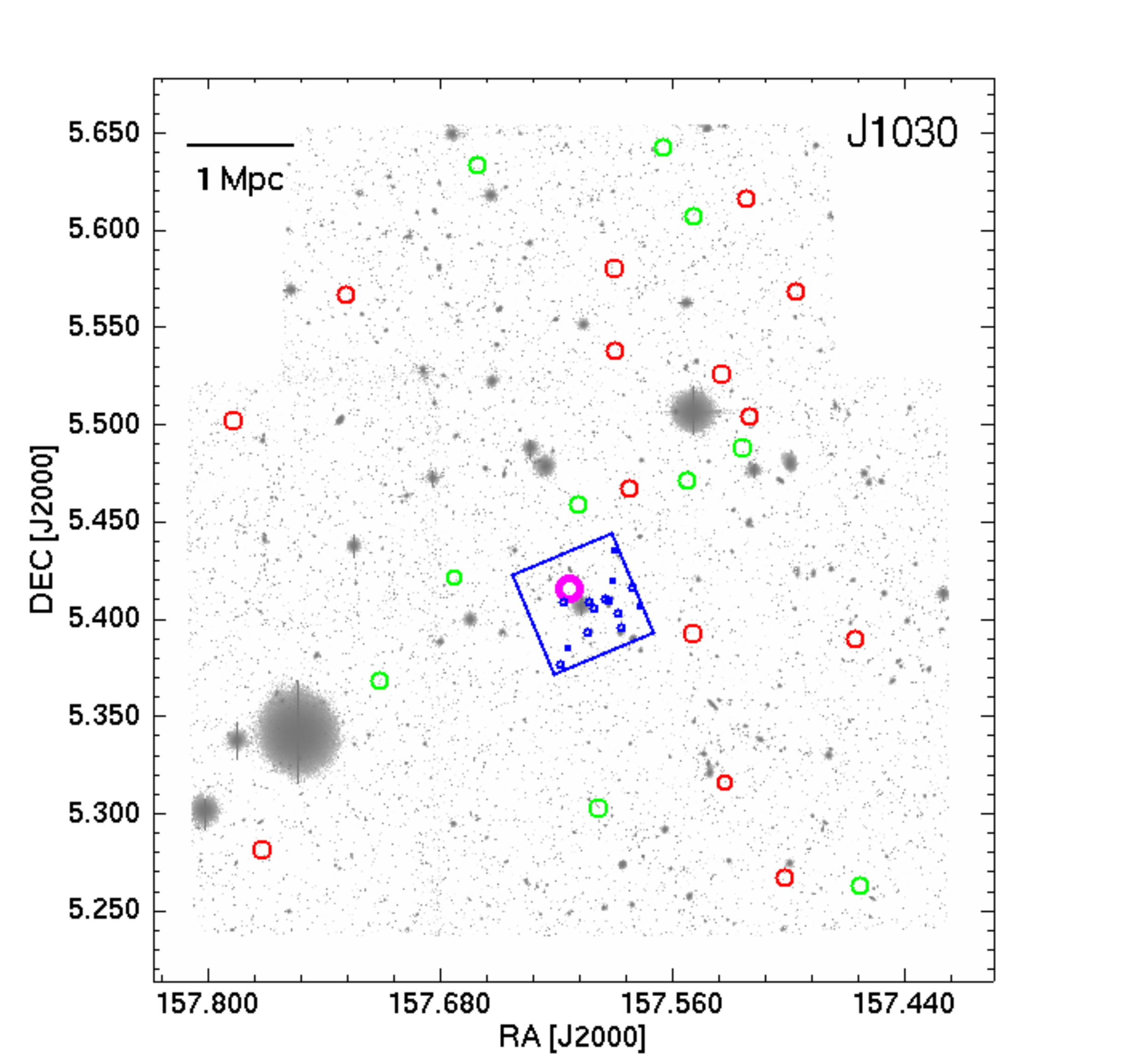}}
  {\includegraphics[scale=0.36]{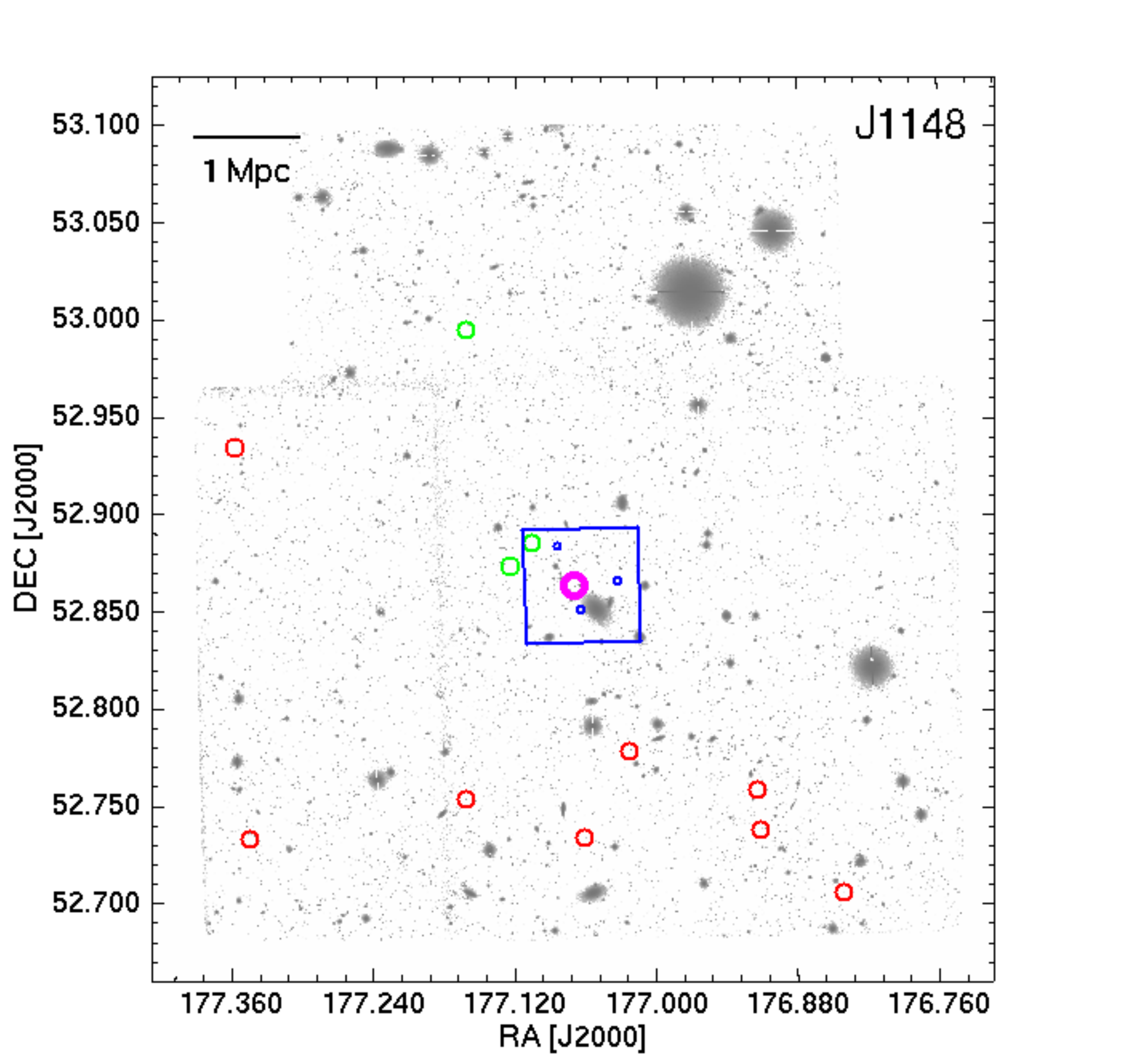}}
   {\includegraphics[scale=0.36]{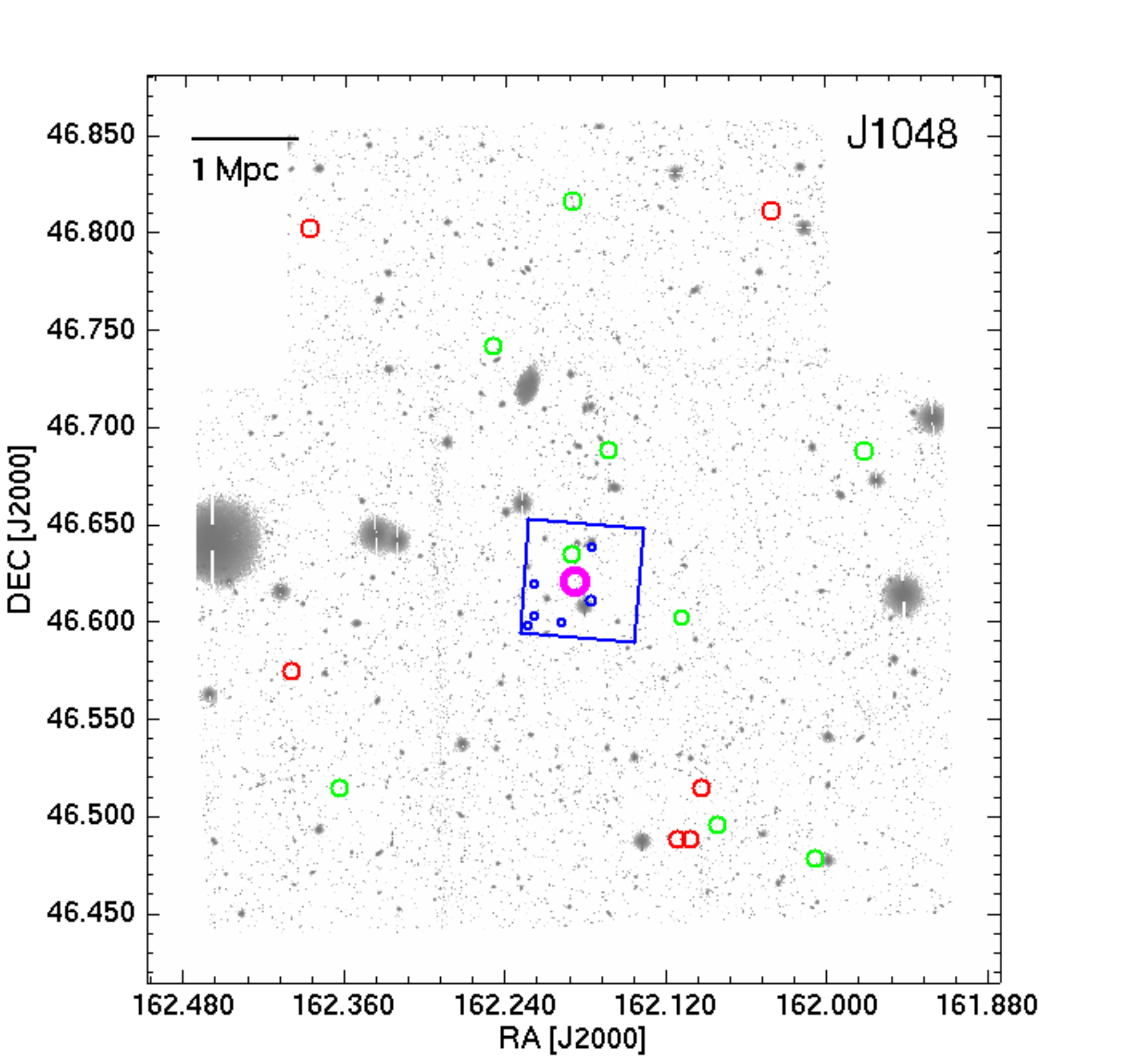}}
   {\includegraphics[scale=0.36]{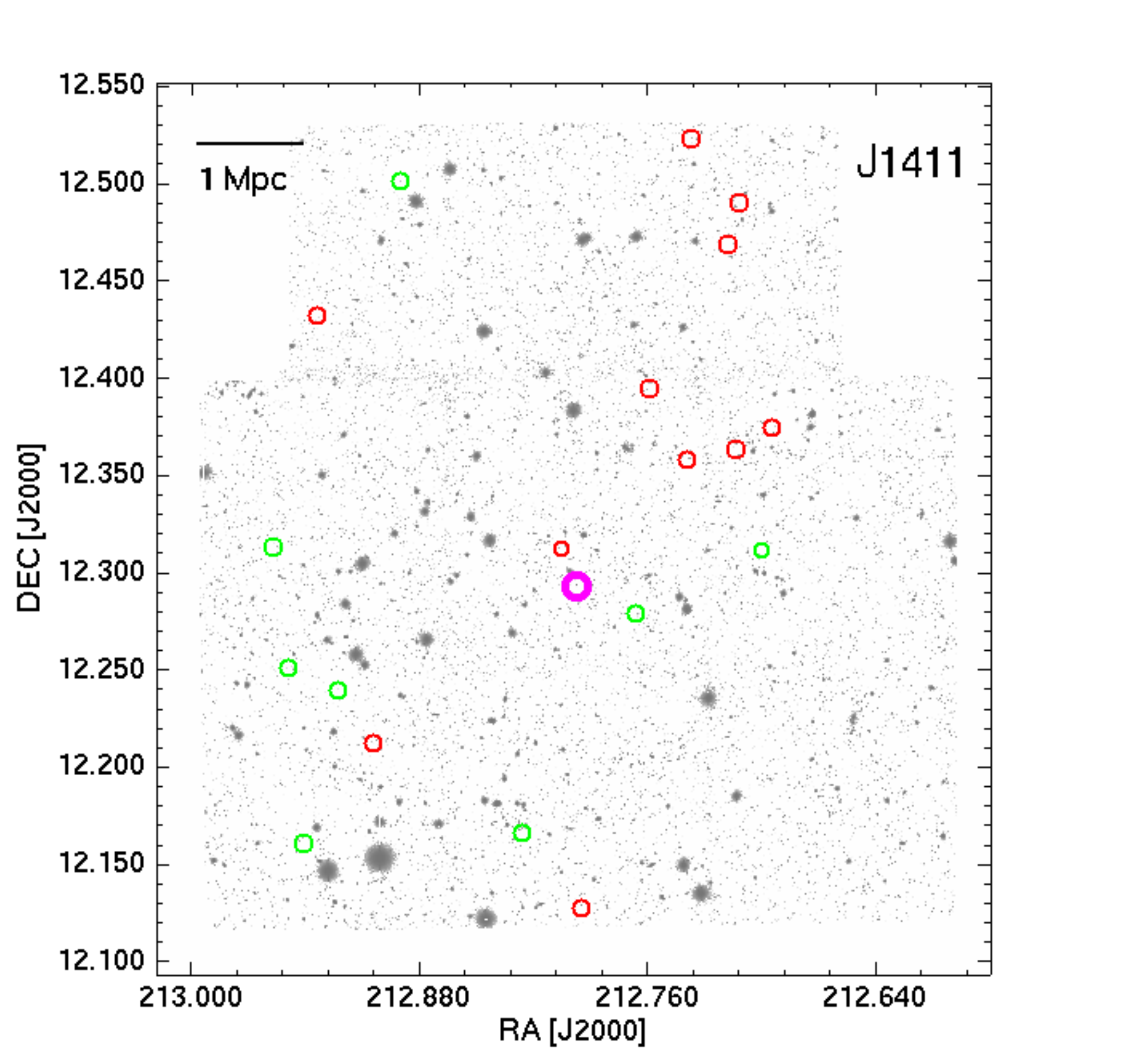}}
%
 \caption{Spatial distribution of $z\sim6$ galaxy candidates around the four QSOs (magenta thick circles) overlaid to the LBC $z$-band images. Red and green circles indicate our primary and secondary candidates, respectively.
The small blue dots mark the faint $i$-band dropouts ($25<z_{850}<26.5$) found in the previous study of K09 carried out with HST/ACS observations. The 3x3 armin$^2$ ACS FoV is shown by the blue squares.}
\label{radec}
 \end{figure*}

Colors have been estimated using the corrected aperture magnitudes, in order to select the flux always from the same portion of the source. Finally, to obtain a more reliable sample of $z\sim6$ candidates, 
we replaced a sharp color criterion with one that includes the photometric color errors. The criteria implemented in this work to select our primary candidates are then as follows:

\begin{itemize}
\item $z_{TOT}$ < $z_{lim, TOT}$;
\item $(i-z) - \sigma_{(i-z)} > 1.3$;
\item undetected in the $r-$band at 3$\sigma$ ($r_{AP}>27.2$);
\end{itemize}

where $z_{TOT}$ is the total magnitude in the $z$ image, $z_{lim,TOT}$ is the completeness limit in $z$, the $i-z$ color and its error $\sigma_{(i-z)}$ are computed with corrected aperture magnitudes,
and $r_{AP}$ is the corrected aperture magnitude in the $r$ image.  

The spectra of high-z QSOs show a strong Lyman $\alpha$ emission line.  At z $\lesssim$ 5.8, the shifted Lyman $\alpha$ line falls in the $i_{SDSS}$ filter. If the line is strong, it causes an enhancement of the flux 
in the $i$ band and, as a consequence, the  $i-z$ colour could be smaller than the adopted threshold of 1.3 (see e.g. \citealt{vanzella10}). 
For this reason, in addition to a primary selection based on $(i-z)-\sigma_{(i-z)}$ > 1.3, we also investigated objects with 1.1 < $(i -z)-\sigma_{(i-z)}$ < 1.3. 
Indeed, it is not uncommon that a significant fraction of objects selected by more relaxed color criteria is made by genuine high-z galaxies \citep{tos}.
Candidates with 1.1 < $(i-z) - \sigma_{(i-z)}$ < 1.3 form our secondary samples and will constitute the additional targets in our future spectroscopic follow-up campaign. 
Fig.\ref{color} shows the position in the $i-z$ versus $r-z$ diagram of the primary and secondary candidates in the four fields. 

We note that the adopted selection criteria cover a rather broad redshift range and may in principle select objects that do not belong to the same structure of the QSO. 
However, the network of LSSs around overdense regions at $z\sim6$ may extend up to 10 physical Mpc in radius \citep{overz}.
This corresponds to $\Delta z\approx0.2$ at $z=6$, and we therefore preferred to avoid more stringent selection criteria that would span only a narrow redshift 
window around each of the four QSOs.

\section{Results}

The number of primary and secondary candidates in the four fields are shown in the last two columns of Table~\ref{riass}, 
while the catalogue ID, $z_{TOT}$, morphological class, $i-z$ and $r-z$ colors of the primary candidates are reported in Table~\ref{primary}. The morphological class has been estimated by analysis of various parameters: 
objects with $z_{TOT}\lesssim24$ are classified as unresolved (point-like; $p$) if CLASS$\textunderscore$STAR (from SExtractor) is greater than 0.5 {\it and} the difference between total and 
corrected aperture magnitude in the $z-$ band is less than 0.1, or as resolved otherwise ($r$). Fainter objects are unclassified ($u$). The spatial distribution of the primary and secondary candidates around the QSOs is shown 
in Fig.\ref{radec}, where the $z\sim6$ candidate sources found in the previous work of K09 are also indicated. The work of K09 was carried out using ACS observations and the $z$-band magnitudes of their candidates 
is in the range [25.0:26.5], i.e. fainter than our completeness limit. We note that $z_{TOT}<25$ corresponds to UV absolute magnitudes of $M_{1450}<-21.7$ at $z=6$, i.e. we are sampling the bright end of the (star forming) galaxy 
luminosity function at that redshift ($\sim0.5$ magnitudes or more brighter than $M^*_{UV}$ at $z=6$, \citealt{bouwens14}). The four fields are characterised by a non-uniform distribution of the candidates  around the central QSO. 
This is an important indicator that our primary samples are probably not dominated by star contaminants. In fact, stars are expected to be randomly distributed in the fields, whereas the distributions of $i-$band dropouts around 
early QSOs are expected to be all but symmetric (see e.g. Fig.~19 of \citealt{overz}).  In J1030, the western region is clearly more populated than the eastern one: 11 out of 14 primary dropouts have RA values smaller than 
the central QSO. Based on the binomial distribution, the probability of observing 11 or more out of 14 objects within half of the field is 2.8\% (1.1\% when adding secondary candidates).
This peculiarity in J1030 was also found by K09 and S05 over a much smaller area around the QSO. In J1148 the southern region is clearly more populated, with 7 out of 8 primary candidates found at lower 
declinations than the QSO (corresponding to a probability of 3.5\% and 0.26\% if the region populated by the 7 southern dropouts is considered to be half or one third of the LBC field, respectively). 
The distribution is less clear in J1048, while in J1411 the majority of the primary candidates is found in the North-West region (the probability of having within one quadrant 7 out of 11 objects is 0.76\%), 
while the secondary sample mostly populates the South-East area. 

\subsection{J1030}
\begin{figure}
  \includegraphics[scale=0.4]{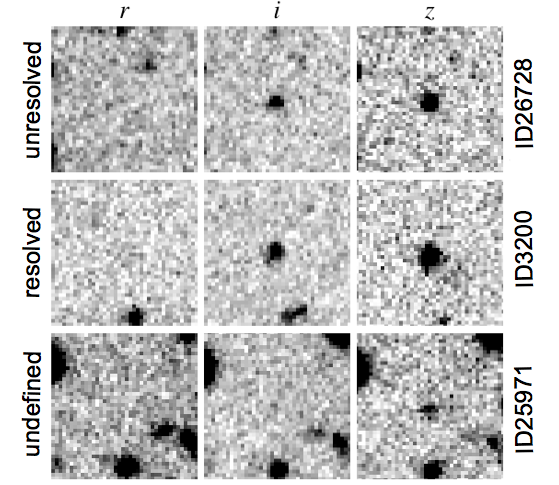}
  \caption{Postage stamps (9 arcsec on a side) in the $z$, $i$ and $r$ filters of three $z\sim6$ LBG candidates in the J1030 field. An unresolved object (top row), a resolved one (center row) and one with unclassified morphology (bottom row) are shown.}
  \label{esempi}
\end{figure}

The field around J1030 hosts 14 $z\sim6$ primary galaxy candidates. Four of them are relatively bright ($z_{TOT}$ < 24.2) and have a reliable morphological classification;  
only one of those four sources is classified as resolved. 
The morphological classification of the remaining 10 fainter sources is not reliable, but a visual inspection of the images and a stacking analysis 
(see Sec. 4.1)  confirm that all of them
are very compact in the sharp $z$ image of the field, supporting the idea that they are good high-z candidates.   
Fig.~\ref{esempi} shows three candidates in the J1030 field in the $z$-, $i$- and $r$- band images: an unresolved, point-like candidate (ID26728), a resolved one (ID3200) and an object with unclassified morphology (ID25971). 
The secondary sample of this field is made of 10 candidates. 

\subsection{J1148}

The primary sample of candidates in the J1148 field is made of 8 sources. The three of them with $z_{TOT}$ < 24  have a reliable morphological classification; only one 
source is classified as an unresolved object in our 
image. This field hosts 3 secondary candidates. \cite{maha} reported the discovery of an optically faint QSO at z = 5.70 at 1.82 arcmin from 
the QSO SDSS~J1148+5251. In our catalogue, 
this QSO is undetected in the $r-$band, has $z_{TOT}=23.4$, and $i-z= 1.11\pm0.05$, and is then just below the selection criteria adopted for our secondary candidates. 
This underlines the importance of studying, in addition to sources with $i-z$ > 1.3 
(our primary sample) also objects with bluer $i-z$ color (part of our secondary sample).

\subsection{J1048}

The field around J1048 is the less populated. It contains only 6  primary candidates. Three of them have $z_{TOT}$ < 24; two have a resolved but compact morphology and one is unresolved.  
The secondary sample is made of 9 sources. 

\subsection{J1411}

The J1411 field is the second most populated field. It contains 11 $z\sim6$ LBG primary candidates, in particular three bright candidates with $z_{TOT} < 24$ (two unresolved and one resolved). 
The secondary sample of $z\sim6$ galaxies around J1411 is made of 8 sources. \\ \\
The $r,i,z$ cutouts of all the primary and secondary dropout samples in the four fields can be consulted at the project webpage.\footnote{\url{http://www.oabo.inaf.it/~LBTz6}}

\setlength{\tabcolsep}{1.8mm}
 \begin{table}
\caption{Object ID, $z_{TOT}$ magnitudes, morphological classification (p: unresolved objects, r: resolved objects, u:undefined), $i-z$ and $r-z$ colors of the primary candidates in the four field. Upper limits on $i-z$ and $r-z$ are indicated.}
\begin{center}
\begin{tabular}{rccccccc}
\hline
 \textbf{ID}&  \textbf{$z_{\textrm{TOT}}$}  & \textbf{Morph}&   \textbf{($i$-$z$)$_{\textrm{AP}}$} $\:$&  \textbf{($r$-$z$)$_{\textrm{AP}}$}  \\
\hline
\hline
J1030\\
\hline
3200	&	23.20$\pm$0.03	&	r		&	1.46$\pm$0.05			&	> 4.76	\\
18619	&	23.51$\pm$0.03	&	p		&	1.36$\pm$0.06			&	> 4.59	\\
26728	&	23.55$\pm$0.03	&	p		&	1.37$\pm$0.06			&	> 4.56	\\
25831	&	24.16$\pm$0.09	&	p		&	1.85$\pm$0.26			&	> 3.39	\\
12265	&	24.22$\pm$0.11	&	u		&	1.58$\pm$0.18			&	> 3.93	\\	
6024	&	24.56$\pm$0.13	&	u		&	1.69$\pm$0.22			&	> 3.44	\\
2140	&	24.75$\pm$0.17	&	u		&	1.62$\pm$0.30			&	> 3.01	\\
21596	&	24.77$\pm$0.11	&	u		&	> 2.01				        &	> 3.34	\\
28941	&	24.87$\pm$0.16	&	u		&	> 1.45				        &	> 2.82	\\
24071	&	24.88$\pm$0.16	&	u		&	1.66$\pm$0.29			&	> 3.19	\\
25971	&	24.92$\pm$0.17	&	u		&	> 1.61				        &	> 2.97	\\
11963	&	25.00$\pm$0.17	&	u		&	> 1.55				&	> 2.91	\\	
23354	&	25.07$\pm$0.14	&	u		&	> 1.51				&	> 2.92	\\
21438	&	25.16$\pm$0.12	&	u		&	> 1.66				&	> 3.01	\\
\hline
\hline
J1148\\
\hline
5113	&	22.99$\pm$0.03	&	p		&	1.50$\pm$0.05		&	> 4.29	\\
3556	&	23.48$\pm$0.07	&	r		&	1.59$\pm$0.16		&	> 2.84	\\
3605	&	23.84$\pm$0.06	&	r		&	1.54$\pm$0.11		&	> 3.33	\\
5393	&	24.04$\pm$0.11	&	u		&	1.52$\pm$0.20		&	> 2.58	\\	
3910	&	24.23$\pm$0.12	&	u		&	1.57$\pm$0.20		&	> 2.66	\\
18368    &	24.68$\pm$0.12	&	u		&	> 2.78 			&	> 2.61	\\
6761	&	24.85$\pm$0.15	&	u		&	> 1.47			&	> 2.30	\\
1556	&	24.91$\pm$0.16	&	u		&	> 1.40			&	> 2.23	\\
\hline
\hline
J1048\\
\hline
3002	&	22.65$\pm$0.02	&	p		&	1.50$\pm$0.03		&	> 4.68		\\
9077	&	23.87$\pm$0.06	&	r		&	1.63$\pm$0.13		&	> 2.73		\\
4811	&	23.93$\pm$0.14	&	r		&	1.83$\pm$0.35		&	> 2.84		\\
22521     &	24.43$\pm$0.12	&	u		&	> 1.65 			&	> 2.47		\\
2968	&	24.57$\pm$0.15	&	u		&	> 1.61			&	> 2.44		\\
22931    &	24.93$\pm$0.16	&	u		&	> 1.54			&	> 2.37		\\
\hline
\hline
J1411\\
\hline
25349	&	22.89$\pm$0.02	&	p		&	1.42$\pm$0.03		&	> 4.42	\\
25638	&	23.48$\pm$0.04	&	p		&	1.50$\pm$0.07		&	> 3.72	\\
564		&	23.75$\pm$0.05  &	r		&	> 2.62			&	> 3.42	\\
19681	&	24.03$\pm$0.06	&	u		&	1.49$\pm$0.10		&	> 3.26	\\
16818	&	24.04$\pm$0.07	&	u		&	1.63$\pm$0.12		&	> 3.27	\\
17304	&	24.29$\pm$0.07	&	u		&	1.79$\pm$0.16		&	> 2.97	\\
6136	&	24.35$\pm$0.11	&	u		&	1.71$\pm$0.17		&	> 2.81	\\
22009	&	24.49$\pm$0.16	&	u		&	> 1.50			&	> 2.33	\\
24189	&	24.84$\pm$0.09	&	u		&	1.59$\pm$0.21		&	> 2.53	\\
18095	&	24.90$\pm$0.17	&	u		&	> 1.42			&	> 2.25	\\
13344	&	24.99$\pm$0.12	&	u		&	> 1.41			&	> 2.24	\\
 \hline
\label{primary}
\end{tabular}
\end{center}
\end{table}

\section{Discussion}

\subsection{Morphology and near-IR colors of primary candidates}
\label{stack}
As explained in Sec.\ref{selection}, the $z\sim6$ candidates samples obtained using a pure color selection may be affected by contamination from elliptical galaxies at lower redshift. 
Since the flux of elliptical galaxies blueward of the $D_{4000}$ does not drop to zero, elliptical galaxies at z $\sim$1-2 are characterised by bluer $r-z$ color than $z\sim6$ sources. 
In addition, the extended structure of ellipticals at z $\sim$ 1-2 should be resolved in the sharp $z$-band images used in this work. This is evident from the 
comparison between the stacking of the 11 primary $z\sim6$ candidates in the J1030 field with $z_{TOT}$ = 24 - 25, for which SExtractor can not give a reliable morphological 
classification, and that of 12 sources in the same range of magnitude, that, albeit being red ($i-z$ > 1.3), do not satisfy our stringent selection constraints, because of their 
larger photometric errors or because they are barely detected in the $r-$filter. We performed this test in the J1030 field since it features the best seeing 
and contains the largest number of $i$-band dropouts (see Table~3). Fig.~\ref{stack} shows the comparison between the profile of the two stacks. 
It is clear that our candidates are unresolved and compact, with a stacked FWHM of the profile 
of 0.6 arcsec, comparable with the profiles of 55 stars in the field, while $i$-band dropouts with bluer $r-z$ color are resolved, with a FWHM of 0.85 arcsec. 
The stacked images of the primary candidates are shown in insets of Fig.\ref{stack}. The $z$-band stacking (left panel) displays a high S/N, unresolved source, that it is not detected on the $r$-band 
stacked image (right panel) with a limiting magnitude of $\approx29$, estimated from the background rms. The unresolved profile in $z$, the non-detection in the $r$-band stacked image, 
and the resulting average color ($\langle r-z \rangle \gtrsim 4$) are all strong clues of the goodness of the adopted color criteria to select $z\sim6$ sources.

In addition to the morphological analysis, near-IR photometry can in principle be used to test the reliability of LBG candidates. For instance,  
$z\sim6$ LBGs are known to be $\sim$1.5-2 mag bluer in $z-J$ than brown dwarfs with similar  $i-z$ color \citep{willott13}. The only deep IR 
observations that cover a sizeable portion of our fields, however, are those by Spitzer/IRAC at $\lambda\geq3.6\mu$m. 
In all fields but J1030, the IRAC coverage is $\sim60$ arcmin$^2$ per field, i.e. $\sim12\%$ of the LBC FoV, and on average only
2 dropouts (mostly secondary) per QSO field have been observed. A large IRAC mosaic is instead available in the J1030 field, covering about half of the LBC FoV, 
for which we retrieved the mosaic images and downloaded the photometry at 3.6 and 4.5$\mu$m from the Spitzer Heritage Archive
\footnote{http://sha.ipac.caltech.edu/applications/Spitzer/SHA} for the 9 dropouts in the mosaics. Although the depth of the IRAC images 
is sufficient to detect most of the dropouts, color-color diagnostics using optical- to-IRAC photometry are far less powerful than 
those using $J$-band data \citep{willott13}. Our dropouts indeed appear to have colors that can be both produced by late type brown dwarfs 
and $z\sim6$ LBGs. A detailed SED analysis of our candidates is beyond the scope of this paper and will be presented elsewhere.

\begin{figure}
  \resizebox{\hsize}{!}{\includegraphics{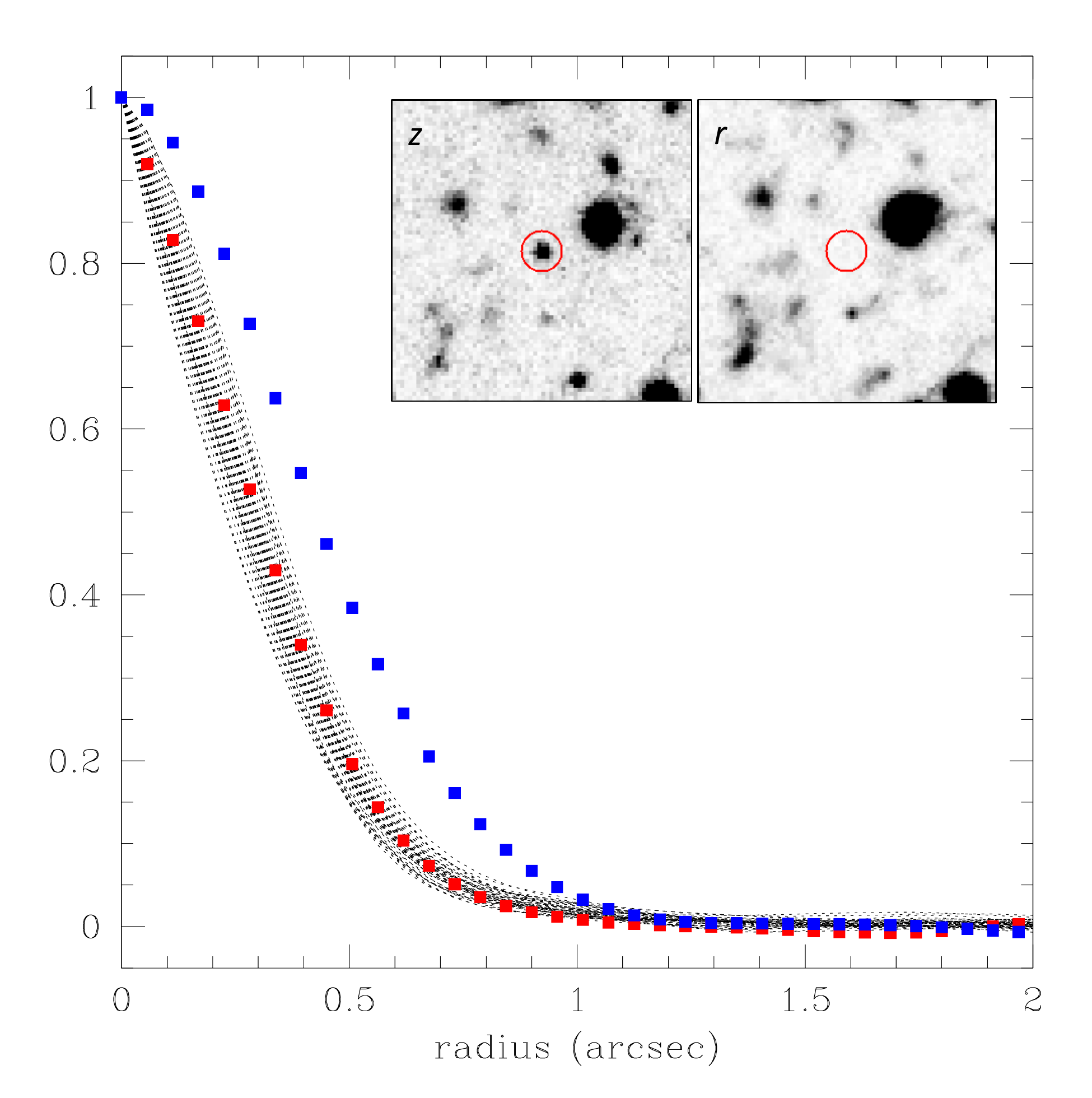}}
  \caption{Profiles of the stacking of the 11 $z\sim6$ primary candidates in the J1030 field with unclassified morphology (red symbols) and of 12 objects with $(i-z)-\sigma_{(i-z)}$ > 1.3 
and detected in the $r$-band image (blue symbols). The dotted lines indicate the profiles of 55 stars selected in the same field to represent the mean seeing (FWHM) of the $z$-band image. 
The insets show the $z$ (left) and $r$ (right) stacked images of the 11 primary candidates; the size of each panel is $15 \times15$~arcsec$^2$.
}
  \label{stack}
\end{figure}


\subsection{Overdensity estimate}

We computed the high-z galaxy overdensity in each LBC field as $\delta = \frac{\rho}{\bar{\rho}} - 1$, where $\rho$ is the number of $i-$band dropouts observed in each 
field and $\bar{\rho}$ is the number of dropouts  expected over a blank sky field of area equal to the LBC FoV (0.144 deg$^2$ after correcting for masked regions). 
To obtain a good estimate of $\bar{\rho}$, a blank sky field as wide as possible is needed, which should also feature observations 
in $r,i,z$ down to a depth similar to, or greater than, that of our LBC observations. To our knowledge, the best available field that satisfies these criteria is the 
Subaru/XMM-Newton Deep Survey  (SXDS; \citealt{fur}), a mosaic of five, partially overlapping Suprime-Cam pointings, covering $\sim1$ deg$^2$ in total, down to B = 28.4, 
V = 27.8, Rc = 27.7, i$'$ = 27.7 and z$'$ = 26.6 (5$\sigma$ detection limit).
The SXDS can be considered a good representation of a typical blank sky field. Its optical source number counts agree well with those of other surveys
in similar bands and with similar depth \citep{fur}. The AGN content of the SXDS is also similar to that of other wide-and-deep X-ray fields. 
The X-ray number counts obtained from moderately-deep XMM observations 
(e.g. $f_{0.5-2}>6\times10^{-16}$ erg cm$^{-2}$ s$^{-1}$) are in good agreement with those from other X-ray surveys, as well as the 
measured clustering level of X-ray sources \citep{ueda08}. Also, the SXDS 
is not biased to any known overdense regions at $z\sim6$. Based on a spectroscopically confirmed samples of Lyman Alpha Emitters
(LAEs), \citet{ouchi08} concluded that the $\sim1$deg$^2$ SXDS has no signature of overdensity or underdensity at $z\sim5.7$, and that the numbers
of LAEs in each of its five subfields are consistent the average field-to-field variations. \citet{mclure06} identified nine robust LBG candidates at $z>5$ in the SXDS, all with $z_{AB}<25$.
One of them was spectroscopically identified by \citet{wil1} as the faintest known QSO at $z\sim6$ (CFHQS J021627-045534, with $z=6.01$, $z_{AB}$ = 24.4, $M_{1450}-22.21$).  
We note, however, that it is not clear whether CFHQS J021627-045534 is an example of {\it bona-fide} high-z QSO. Its spectrum is indeed very different from that commonly observed
for $z\sim6$ QSOs, since it features a single prominent Ly$\alpha$ line without any continuum emission. Its spectrum thus resembles those of bright LAEs, and its absolute magnitude 
is also consistent with those observed at the bright end of the $z\sim 6$ galaxy luminosity function \citep{willott13,bouwens14}. Furthermore, the relatively large FWHM measured by \citet{wil1}
($\sim1600$ km s$^{-1}$) may be ascribed to gas outflows produced by star formation. 
At any rate, even assuming that CFHQS J021627-045534 is powered by an accreting BH, its mass would be smaller than that of the bright QSOs analyzed in this work by about two dex 
(assuming it is accreting at the Eddington limit, like most $z\sim6$ QSOs do). Therefore, it would likely trace more typical, less dense environments at $z\sim6$.

Five $z$-band selected catalogs, one for each pointing of the Subaru Suprime-Cam are publicly available for the SXDS
\footnote{\sl http://soaps.nao.ac.jp/SXDS/Public/DR1/index$\textunderscore$dr1.html}.
We retrieved and merged them to create a single master catalogue, removing multiple measurements of the same object in the overlapping regions. 
The effective area of the resulting SXDS master catalog is $\sim$1.13 deg$^2$ (see also Table 3 of \citealt{fur}).
To compare our LBC catalogues and the SXDS one, a correction factor that takes into account the offset between the different filters has been computed, comparing the photometric 
measurements of SXDS to those of the  SDSS in the same area of the sky\footnote{The average values of psfMag(SDSS) - MagTot(SXDS) are -0.05,-0.03, and 0.11 in $z,i,r$, respectively. 
This leads to: MagTot(LBC) - MagTot(SXDS) = 0.05, -0.1, 0.28 in $z, i, r$, respectively.}.

\begin{figure}
  \resizebox{\hsize}{!}{\includegraphics{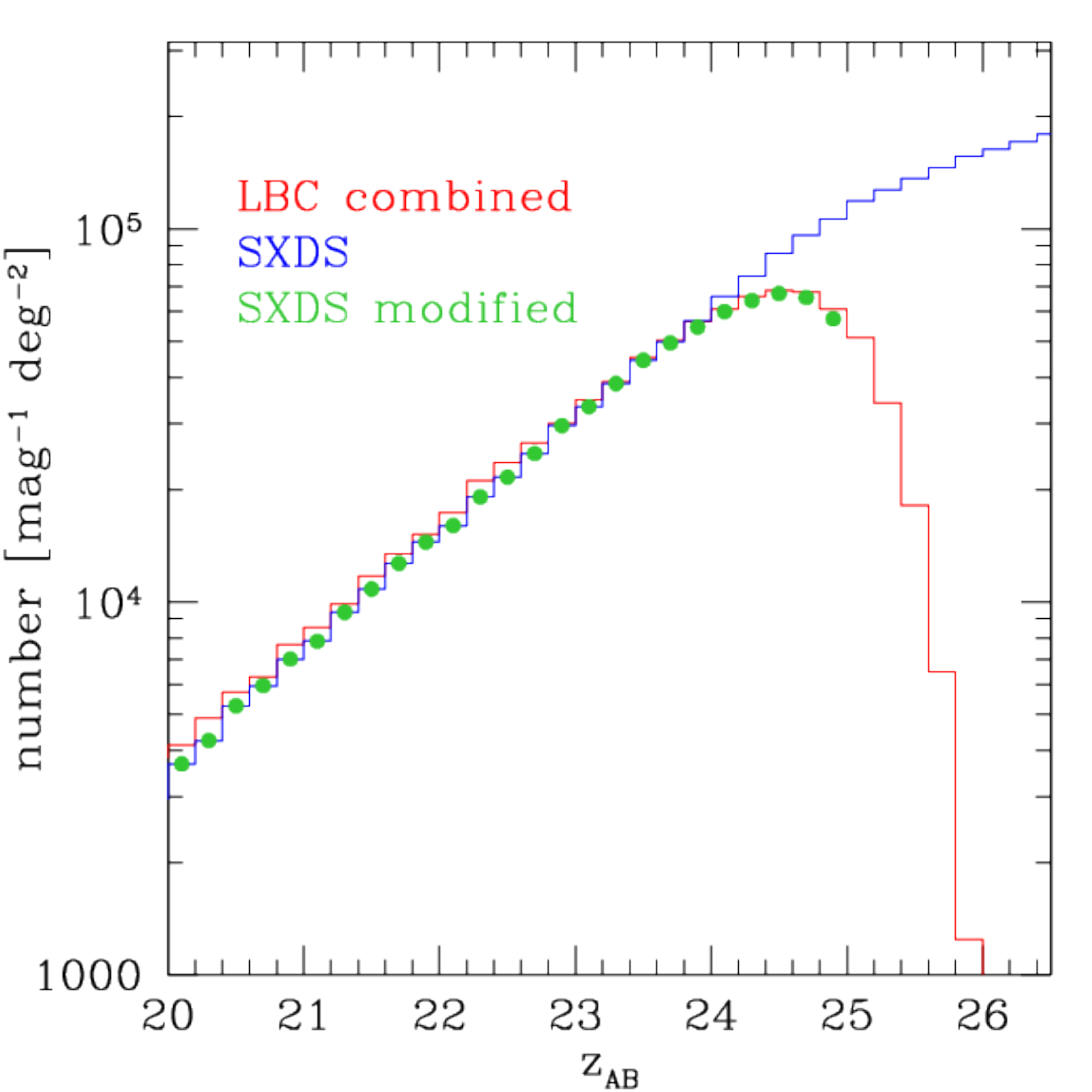}}
  \caption{Source number counts in the $z$-band for the LBC combined catalog (red histogram), the SXDS catalog (blue histogram) and the SXDS modified catalog (green points).
 The SXDS modified catalog, cut here at $z_{TOT}<25$, represents the average of the 10000 simulated SXDS catalogues downgraded to the completeness level of the LBC fields (see text for details).}
  \label{compl}
\end{figure}

Our catalogues and the SXDS one have different completeness limits and then different source densities. In fact, the LBC catalogues start missing objects at 
$z_\textrm{TOT}\approx 23.8$ , whereas the SXDS one is still 100$\%$ complete. Most LBG candidates are in the range 24 < $z_\textrm{{TOT}}$ < 25. Therefore, if a correction to this 
effect is not added, the overdensity would be underestimated. To solve this problem, 10000 SXDS catalogues have been simulated, differentially excluding objects from the SXDS catalogue 
in the range 24.0 < $z_\textrm{{TOT}}$ < 25.0 in such a way that the simulated catalogues reproduce the same trend of the number of counts with z-magnitude of the LBC observations
(Fig.~\ref{compl}). It is clear that the average of the SXDS modified catalogues has the same drop of 
counts of the LBC fields near the completeness limit (cut at $z_\textrm{{TOT}}$ = 25.0).  We remark the good agreement between the number counts in our LBC combined catalog and those in the SXDS at $z_{TOT}<23.8$,
and note that a similarly good match between our LBC fields and the SXDS is also found when comparing the source counts in the $r$ and $i$ band, as well as the $i-z$ vs $r-z$ color-color diagrams. This
makes us confident that the corrections for the different photometric systems have been done properly and hence the SXDS can be used as a good comparison field to compute galaxy overdensities. 

We then verified whether the dropout selection criteria presented in Section 2.3 can be safely applied to select $z\sim6$ galaxy candidates in the SXDS. The errors on the magnitudes of the SXDS galaxies 
as reported by \cite{fur} appear to be one order of magnitude smaller than what we find for LBC sources of comparable magnitudes. We performed an independent estimate of the magnitude errors of 
SXDS galaxies by considering those objects that have been observed twice by Suprime-Cam and for which two independent magnitude estimates are then available. We computed the magnitude 
differences ($\Delta mag$) for these objects in five magnitude bins: the rms of the $\Delta mag$ distributions is significantly larger than the errors quoted in the SXDS catalog at the same 
magnitudes.
Therefore, since we do not have full control of the relative errors between our LBC fields and the SXDS, we preferred to demise any dropout selection criteria which involves errors on magnitudes. 
We instead adopted a simple (and more conservative) color cut at $i-z$ > 1.4 to define the $i$-band dropout samples to be used for the overdensity estimate. 
Table~4 shows the number of candidates in the LBC fields that satisfy these criteria. However, because the photometric errors in our LBC catalogs are larger than those in the SXDS, it is likely that our 
dropout samples suffer from more contamination from objects with $i-z<1.4$, which are more abundant than those with $i-z>1.4$, and hence can artificially enhance the observed number of the latter 
if photometric errors are significant. We tried to quantify this contamination as follows: we assumed that the $i-z$ distribution of SXDS objects with $z_{TOT}<25$ and $r_{AP}>27.2$ is the true, underlying 
$i-z$ distribution of our LBC dropout candidates (note that this is conservative, since it assumes that the SXDS catalog has negligible photometric errors). Then we randomly extracted an $i-z$ color from 
this input distribution, assumed a Gaussian error with $\sigma=0.2$ on it (i.e. the average error on the $i-z$ color measured for our LBC dropout samples; see Table 1), and extracted a new $i-z$ color from the 
Gaussian distribution. We repeated this process 10000 times and built a new $i-z$ distribution that is indeed broader than the input one, because of photometric errors. We compared the input and 
the ``blurred'' $i-z$ distributions and verified that the fraction of objects with $i-z>1.4$ has increased by 20\% from the input to the blurred one. Therefore, we estimate that 20\% of our observed 
dropouts with $i-z>1.4$ are bluer objects that met the selection because of photometric errors, and hence quote in Table~4 also the number of ``statistically decontaminated'' dropouts in the four LBC fields.

\begin{figure}
\centering
\includegraphics[scale=0.4]{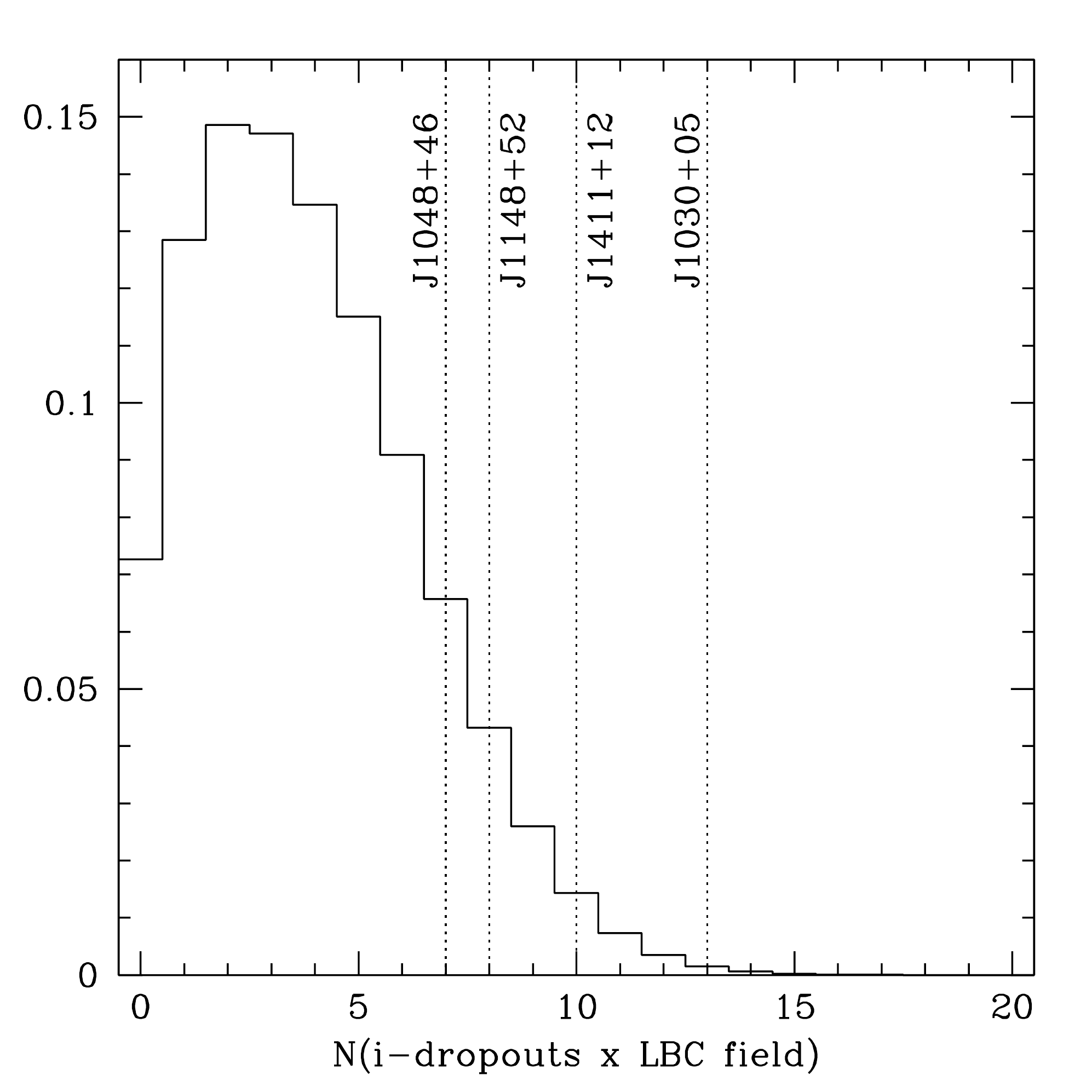}
\renewcommand{\figurename}{Fig.}
\caption{Number of sources measured in 10000  Poisson realisations based on the completeness-corrected number of dropouts detected in 23x25 arcmin$^2$ rectangles randomly placed within the SXDS field. 
The vertical lines represent the number of ``decontaminated'' dropouts with $i-z>1.4$ in the four LBC fields (see Table 4 and text for details). }
\label{poiss}
\end{figure}

We then estimated the number of dropouts that has to be expected in a 0.144 deg$^2$ area based on the SXDS catalog. We applied the same selection criteria as above ($z_{TOT}<25$, $r_{AP}>27.2$, $i-z>1.4$) 
to the SXDS considering only those objects in the 0.95 deg$^2$ of the SDXS that are free from either strong or weak halos caused by bright stars, and whose photometry should be then more reliable.
These candidates were visually inspected to remove clear artifacts and spurious sources, exactly in the same way we cleaned our LBC catalogs. We finally obtained a clean sample of 41 SXDS dropouts 
over 0.95 deg$^2$.  Based on the 10000 SXDS modified catalogues, this number has to be reduced by $\sim30\%$ to account for the incompleteness of the LBC fields,
which translates into $4.3$ sources expected per LBC FoV. Based on the number of ``decontaminated'' LBC dropouts presented in Table 4, we then measure an overdensity of 2.0, 0.9, 0.6, 1.3 in 
J1030, J1148, J1040, J1411, respectively. The total overdensity that is obtained by summing the four fields is 1.2. \\

To correctly estimate the statistical significance of the overdensity measured in the four LBC fields, the effects of cosmic variance have to be taken into account.
Once again we exploited the SXDS to evaluate it, randomly placing within the SXDS area rectangles of 25x23 arcmin$^2$, i.e. the LBC FoV, counting the number of dropouts in each rectangle and correcting 
for incompleteness. An average of four (completeness corrected) dropouts per rectangle is measured, and no rectangle contains more than 8 dropouts. For each rectangle, we then built a Poissonian distribution 
with mean equal to the completeness-corrected number of dropouts, and summed all the Poissonian distributions
to build the total distribution presented in Fig.~\ref{poiss}, which then accounts for both statistical uncertainties and cosmic variance (albeit limited to what can be extracted from the SXDS field). 
The probability of observing in this distribution a number of dropouts equal to or larger than what is observed in J1030, J1148, J1048, J1411 is 
$1.0\times10^{-3}$, $5.4\times10^{-2}$, $9.7\times10^{-2}$, $1.3\times10^{-2}$, respectively. If the distribution were Gaussian, these 
probabilities would correspond to 3.3, 1.9, 1.7, 2.5$\sigma$, respectively. The two fields J1030 and J1411 are the densest, with an overdensity significance of $>2\sigma$. 
J1148 and J1048 are also denser than average, but with lower significance.

As discussed at the end of Section 2.3, the adopted i-z color criteria cover a broad redshift range that may in principle select objects not associated to the QSO LSS.
We then performed a test on the J1030 field, where the dropout statistics is the highest, by selecting i-band dropouts with $i-z>1.8$, that are expected to be at $z>5.9$ \citep{beck}, i.e. within
$\sim20$ physical Mpc from the QSO (z=6.28). By repeating the same analysis described above, we found that the observed overdensity in J1030 is still significant at the $\sim2.5\sigma$ level.

We finally considered what is the probability of finding a number of dropouts equal to or greater than what is observed in the four LBC fields all together, 
by randomly extracting four of the LBC-like rectangles at a time. We found that $z\sim 6$ QSOs are surrounded by galaxy overdensities at the 3.7$\sigma$ level.\\

\begin{figure}
  \resizebox{\hsize}{!}{\includegraphics{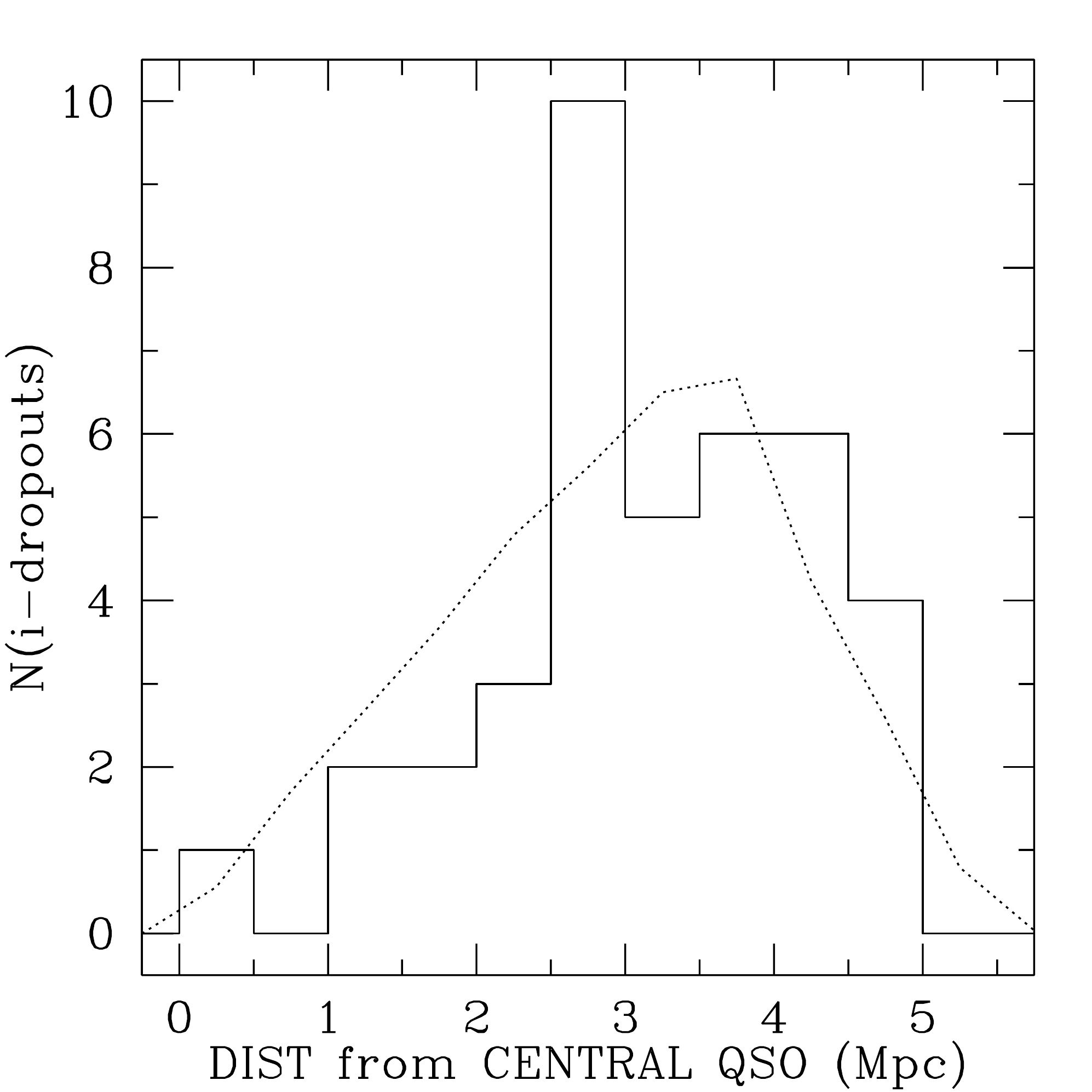}}
  \caption{
Distribution of the primary candidate projected distances (in physical Mpc) from the central QSO, in all the four fields (solid histogram). 
The dotted curve represents the distribution of the $\sim75000$ objects with $22<z_{TOT}<25$ in the LBC photometric catalogs, normalized to the total number of dropouts (39). 
The drop in the galaxy density at radii larger than $\sim$4 Mpc is due the physical limit of the LBC FoV.
}
  \label{radist}
\end{figure}

\subsection{Spatial distribution and comparison with previous results}


In three of our LBC fields (J1030, J1148, J1048) the density of faint ($27<z_{850}<25$) $i-$band dropouts 
was already investigated through observations with HST/ACS on scales a factor of $8$ smaller than those sampled in this work (S05, K09).
In J1030, both S05 and K09 observed an overdensity with respect to the reference fields (GOODS South and North). The significance of the overdensity obtained in our work, simply based on the dropout counts alone,
is larger than what was measured by K09 and S05 ($2.8\sigma$) on the same basis (S05 and K09 report a higher significance when the dropout color distribution is also taken into account).
 K09 also found that J1148 and J1048 have a dropout density lower than and similar to that of the field, respectively, whereas we find that, at large scales, 
both fields are denser than average (albeit with low significance). Although it is difficult to make a direct comparison between the dropout densities observed by K09 and those reported in this
paper, since they sample different magnitude regimes and have been also selected with slightly different color criteria, it is worth making a few considerations: in J1030, the most overdense field
in both K09 and our sample, the distribution of i-band dropouts is strongly skewed westward of the QSOs in both ACS and LBC data, which then are fully consistent with each other.
In J1148, we found that most of the LBC dropouts are distributed south of the QSO, at separations from the QSO larger than 5 arcmin. Therefore, if the LBC dropouts are tracing the most overdense galaxy region
in the J1148 field, this would explain why the ACS observations have missed it. 
In general, previous results based on narrow-field ($<5\times5$ arcmin$^2$) imaging around high redshift QSOs have produced contradictory results in terms of galaxy overdensity. 
Our results, coupled to those of Utsumi et al. (2010) who measured a $\sim3\sigma$ overdensity of $i-$band dropouts around a $z=6.43$ QSOs using the wide-field Subaru/Suprime-Cam 
(34$\times$27 arcmin$^2$ FoV), suggest instead that large-scale galaxy overdensity are common among high-z SMBHs. 

In Figure~\ref{radist} we compare the distribution of the projected distances from the central QSO (in physical Mpc) of our primary candidates,
with the distribution of the $\sim75000$ objects with $22<z_{TOT}<25$ detected in our LBC fields. The i-band dropout distribution suggests a deficit of high-z galaxies at $r<2.5$Mpc 
when compared with the random field galaxy distribution. A K-S test shows that the radial distribution
of high-z candidates is different from that of the field galaxies at a 98.5\% confidence level (2.4$\sigma$). A similar deficit was also reported by \citet{ut} on the same scales.
This could be an indication that, on scales of 2-3 physical Mpc around the QSOs, i.e. comparable with their Stromgren spheres \citep{kurk,wyithe}, 
galaxy formation is suppressed because of the strong QSO negative feedback (either radiative or even mechanical). Such an interpretation, among other possibilities, was put forward by \cite{romdiaz} to explain
why the average number of dropouts per ACS field measured by K09 (4 per field) was a factor of $\sim2$ lower than that predicted by their simulations that placed early BHs in halos with $M>10^{12}\;M_{\odot}$ 
surrounded by overdense environments. The other obvious interpretation  is that $z\sim6$~QSOs are not hosted in the most massive collapsed halos (see e.g. Fanidakis et al. 2013). 
However, if our results showing large scale overdensities in all the four QSO fields will be confirmed by optical spectroscopy, as well as deeper observations, then the negative-feedback interpretation 
would be favored, and it will be demonstrated that early SMBHs do indeed live in the densest environments at their time.

\setlength{\tabcolsep}{2mm}
 \begin{table}
\begin{center}
\begin{tabular}{ccccc}
\hline
\hline
Field& $\rho$&  $\rho_d$& $\delta$& $\sigma_{\delta}$\\
(1)&  (2)& (3)& (4)& (5)\\
\hline
J1030&	16&	  13	& 2.0& 3.3\\
J1148&	10&	    8	& 0.9& 1.9\\
J1048& 	  9&	    7	& 0.6& 1.7\\
J1411&	12&	  10	& 1.3& 2.5\\
\hline
\label{conserv}
\end{tabular}
\caption{Summary of measured overdensities. (1) QSO field. (2) Number of observed dropouts defined as objects with $z_{TOT}<25$, $r_{AP}>27.2$, $i-z>1.4$. (3) Same as Column 2 but corrected for the contamination from
bluer objects caused by photometric errors (see text for details): these are the numbers adopted to compute the overdensities and their significance, shown in Columns 4 and 5, respectively. 
(4) Overdensity  $\delta = \frac{\rho}{\bar{\rho}} -1$, where $\bar{\rho}=4.3$, as estimated from the SXDS. (5) Significance of the measured overdensities (expressed in Gaussian-equivalent probabilities; see text for details). }
\end{center}
\end{table}

\section{Conclusions}

We have studied large-scale galaxy overdensities around four bright QSOs hosting four of the most massive BHs at $z\sim6$ known to date 
(namely SDSS J1030+0524, SDSS J1148+5251, SDSS J1048+4637, and SDSS J1411+1217) 
using deep  images in $r$, $i$ and $z$ obtained with the 23x25 arcmin$^2$ Large Binocular Camera (LBC) at the Large Binocular Telescope (LBT). As opposed to most previous studies that were sampling relatively small scales, 
our observations map a region of $\sim8\times8$ physical Mpc at $z\sim6$, and hence can trace galaxy overdensities on scales comparable to those expected by simulations of early structure formation. 
We built z-band selected catalogs down to $z_{AB}=25$, performed i- and r-band photometry, and applied different color-color selection techniques to identify galaxies at approximatively the same redshift of the four QSOs.
Our main results are as follows:\\

\begin{itemize}
\item
The four analysed fields show a significant number of candidate $z\sim6$ galaxies selected as $i-$band dropouts. Their spatial distribution in the LBC fields is highly asymmetric, as expected by simulations of 
galaxy formation in high-z dense environments. This argues against a major contamination from late-type stars, which should be more evenly distributed within the fields.
A possible deficit of dropouts within 2.5 physical Mpc from the QSOs is found at a 2.4$\sigma$ level. This suggests that feedback from the QSO may hinder galaxy formation 
within those scales.
\item 
Most of the $i-$band dropouts have $z_{TOT}>24$ and no reliable morphological information can be derived for them. However, the stacking of these faint dropouts in the J1030 field, which features 
the sharpest and deepest observations, shows that they are unresolved and compact, with a FWHM of the median profile comparable with the seeing of the $z$-band image. 
This excludes severe contamination from lower-redshift early-type galaxies and reinforces the likelihood of having selected genuine high-z objects.
\item
The number of $i-$band dropouts measured in our fields is always larger than what would be expected in a blank sky field. By considering the Subaru/XMM Deep Survey (SXDS) as a reference blank field,
and after accounting for the different LBC vs SXDS depths and photometric systems, we expect only 4.3 $i-$band dropouts 
(defined as those objects with $z_{TOT}<25$, $r_{AP}>27.2$, $i-z>1.4$)  over a 0.14 deg$^2$ blank field, whereas 16, 10, 9, 12 dropouts were observed in the J1030,  J1148, J1048 and J1411 field, respectively. 
When accounting for cosmic variance and for the contamination by bluer objects produced by photometric errors, these numbers correspond to overdensity significancies of 3.3, 1.9, 1.7, 2.5$\sigma$, respectively. 
By considering the number of dropouts in the four LBC fields all together and comparing it with what is expected in four random blank fields of 0.14 deg$^2$ each, we find that high-z QSOs reside in overdense 
environments at the $3.7\sigma$ level.
\item
Three out of four QSO fields were already observed by the 3x3 arcmin HST/ACS. The ACS observations revealed the presence of an $i-$band dropout overdensity only in the J1030 field, whereas the J1148 and J1048 fields 
were found to be under-dense and average-dense, respectively. The overdensity that we measure in each of the four fields (albeit the significance is below 2$\sigma$ in two of them), suggests that wide-field observations
may be more effective in revealing large-scale structures around early QSOs. This would also be consistent with the idea that, although early SMBHs reside in dense environments, 
galaxy formation may be depressed in their immediate vicinity because of their strong radiation field.
\end{itemize}

Follow-up spectroscopy of the selected i-band dropouts is obviously needed to confirm whether they are mostly made by genuine $z\sim6$ galaxies. Also, the extension of our LBC imaging analysis to other
QSO fields would be highly desirable to get a statistically sound sample. We started the process of getting optical spectroscopy, as well as near-IR imaging on some of our fields to further refine the 
selection of $z\sim6$ galaxies. Other multi-band data, including X-rays, might reveal the presence of AGN within the observed structures at $z\sim6$ and would place the first observational constraints to
the environment of primordial objects, thus providing a fundamental input to models of early BH and galaxy formation.

\begin{acknowledgements}
We acknowledge the support from the LBT-Italian Coordination Facility for the execution of observations, data distribution and reduction.
We acknowledge support from the Italian Space Agency under the ASI-INAF contract I/009/10/0 and from INAF under the contract PRIN-INAF-2012.
MB acknowledges support from the FP7 Career Integration Grant ``eEASy'' (``SMBH evolution through cosmic time: from current surveys to eROSITA-Euclid AGN Synergies", CIG 321913).
The referee is acknowledged for insightful comments that improved the quality of the paper.
\end{acknowledgements}


\bibliography{ref}
\bibliographystyle{aa}

\end{document}